\documentclass{aa}  
\usepackage{graphicx}
\usepackage{txfonts}
\begin{document}
   \title{Power spectrum Analysis of Far-IR Background Fluctuations 
in Spitzer maps at 160 $\mu$m}

   \author{B. Grossan
          \inst{1,2}
          \and
          G. F. Smoot\inst{3}
               }
          
    \institute{Eureka Scientific, Inc.,
    	2452 Delmer Street Suite 100, Oakland, CA 94602-3017,
	\and
	 Institute for Nuclear and Particle Astrophysics, 50R-5008, Lawrence Berkeley National Laboratory, 1 Cyclotron Road, Berkeley, CA 94720-8158\\
              \email{Bruce\_Grossan@lbl.gov}
         \and
	University of California at Berkeley, Department of Physics, LeConte Hall, Berkeley, CA 94720\\
	             \email{GFSmoot@lbl.gov}
                  }

\offprints{B. Grossan}
 
   \date{2006 April 17 (S11 data, misc. improvments, figs not integrated, i.e. submission text but not graphics.)}

 
  \abstract{We describe data reduction and analysis of fluctuations in the Cosmic Far-IR Background (CFIB) in large maps observed with the Multiband Imaging Photometer for Spitzer (MIPS) instrument 160 ${\mu}m$ detectors.  We analyzed the extragalactic First Look Survey (FLS) and the Spitzer Wide-area Infrared Extragalactic Survey (SWIRE) Lockman Hole observations, the latter being the largest low-cirrus mapping observation available.  In the Lockman Hole map, we measured the power spectrum of the CFIB by fitting a power law to the IR cirrus component, the dominant foreground contaminant, and subtracting this cirrus signal.  The CFIB power 
spectrum at mid -high \textit{k} (\textit{k} \ensuremath{\sim} (2 - 5) \ensuremath{\times} 10$^{-1}$ 
arc min$^{-1}$) is consistent with previous measurements of a relatively 
flat component. At lower \textit{k}, however, the power spectrum is 
clearly not flat, decreasing from our lowest frequencies (\textit{k} 
\ensuremath{\sim} 3 \ensuremath{\times} 10$^{-2}$ arc min$^{-1}$) and flattening at mid-frequencies 
(\textit{k} \ensuremath{\geq} (1 - 2) \ensuremath{\times} 10$^{-1}$ arc min$^{-1}$). This 
behavior is consistent with the gross characteristics of predictions 
of a source clustering ``signature'' in CFIB power spectra, and this is the first report of such a detection.
}  
               

  \maketitle
 
%

\section{Introduction}


Spatial fluctuations in the cosmic far-IR background (CFIB) were 
first discovered with ISO at 170 \ensuremath{\mu}m in a relatively small 
field (0.25 deg$^{2}$; Lagache \& Puget  (\cite{LagachePuget00}). They have also been 
detected with IRAS data, after re-processing (Miville-Deschenes, 
Lagache \& Puget \cite{Deschenes02}). The origin of the CFIB is believed to 
be the ensemble emission from galaxies too faint to be resolved; 
the spectrum, intensity, and spatial distribution of the CFIB 
therefore contain information about the distribution of galaxy 
emission in space and time. Lagache, Dole, and Puget (\cite{LDP03}) showed 
that while a variety of luminosity functions might fit far-IR 
source counts and total background level, the fluctuations in 
the CFIB require that the luminosity function must change dramatically 
with redshift, with a rapid evolution of the high-luminosity 
sources (L $>$ 3\ensuremath{\times}10$^{11}$ L$_{sol}$) 
between z $= $ 0 and 1. 

CFIB fluctuations should also yield information on the spatial 
distribution, i.e. clustering, of IR emitting galaxies. A  
Gaussian distributed field of sources would yield a 
flat CFIB power spectrum.  Perrotta et al. (\cite{Perrotta}), however, have predicted that 
a large excess would be present in the low- to mid- frequencies of 
the power spectrum if source clustering is present. Lagache \& Puget  
(\cite{LagachePuget00}) states that it was not possible to measure a 
clustering signal in the data analyzed, primarily because the small field size
 caused difficulties in fitting and separating the cirrus signal.

\begin{table*}
\caption{Map Fields}
\label{table_fields}
\centering
\renewcommand{\footnoterule}{}  
\begin{tabular}{cccccccc} 
\hline \hline
{Field} & {RA\footnote{a} } &  {Dec$^1$}  & {Area}  & {Square Area\footnote{b} }& {ISM Bgnd} &  {Total Bgnd} & {Total Bgnd}\\
         &   (J2000)      &    (J2000)        &             &                         & (Predicted)           & (Predicted)              & (Measured) \\
 &  &  & (deg$^{2}$)  & (deg$^{2}$) & (MJy/Sr)  &   (MJy/Sr)  & (MJy/Sr)   \\
\hline
FLS& 
 17:18:00  & 
 +59:30:00 & 
 4 & 
 4.02 & 
 2.4 & 
 4.4 & 
 5.86\\

SLH& 
10:47:00 & 
58:02:00 & 
 14.4 & 
 8.47 & 
 1.1 & 
 3.6 & 
 4.27 \\
 \hline
\end{tabular}  
\end{table*}



The Spitzer Space Telescope MIPS instrument (Rieke et al. \cite{Rieke}) 
has observed much larger fields than Lagache \& Puget  (\cite{LagachePuget00}) with 
low IR cirrus emission that are 
ideal for study of the structure of the CFIB. The longest wavelength 
array has a nominal bandpass of \ensuremath{\sim}35 \ensuremath{\mu}m around 
its center wavelength of 160 \ensuremath{\mu}m, and is the most sensitive 
instrument to date in this wavelength range. Cirrus and zodiacal 
emission are weak relative to the unresolved CFIB 
in this band. At 160 \ensuremath{\mu}m,  18\% of the CFIB light is
predicted to be resolved into sources with 
MIPS (Dole, Lagache, \& Puget 2003) so CFIB emission actually 
 dominates source emission in this band. 

In order to take advantage of the capabilites of MIPS for CFIB 
fluctuation studies at the longest wavelength, we have undertaken 
a program to reduce and analyze the largest low-background fields 
observed by the MIPS instrument. This paper describes the current 
status of our efforts to construct and analyze these large 160 
\ensuremath{\mu}m Spitzer MIPS maps using power spectrum analysis. The 
basic characteristics of the map observations that we analyzed 
are given in Table 1.

\section*{2. Map Observations and Reduction}

\subsection*{2.1 Observations}

As of this writing, the largest contiguous low-cirrus field with 
good coverage is the SWIRE Lockman Hole field (hereafter SLH). 
We also reduce the First Look Survey (FLS) extragalactic field, 
the first released, for comparison. As can be seen in Table I, 
the SLH field is considerably lower in cirrus background (as 
known from IRAS and HI maps prior to observations) and larger. 


The maps were made in scanning observation mode, with the MIPS 
arrays performing simple back-and-forth scans. At the end of 
each scan (except in a small fraction of the data), the pointing was stepped by (148"/276") 
in the cross-scan direction in the SLH/FLS surveys (nominally 
just under half the 160 \ensuremath{\mu}m array width , \ensuremath{\sim}9.25 
pix, for the SLH/ nominally 85\% of the 160 \ensuremath{\mu}m detector 
width , \ensuremath{\sim}17.25 pix for the FLS). All SLH scanning observation 
sequences, or AORs, were performed twice back-to-back. (The basic 
unit of planned observation activity with Spitzer is an AOR, 
or Astronomical Observing Request. Here we refer to the series 
of actions by the observatory, and the associated data, as the 
AOR for brevity.) A simple representation of the scan pattern 
is given in Figure 1 for both maps.

\subsection*{2.2 Instrument Behavior - Challenges with Ge detectors}


The far-IR instruments on Spitzer, the MIPS 70 and 160 \ensuremath{\mu}m 
cameras, use arrays of Ge:Ga and stressed Ge:Ga detectors, respectively. 
Ge detectors are subject to random and 1/f noise components, 
including gain drift, and ``memory'' effects which 
are extremely difficult to model and correct. (The so-called 
memory effects are dependences in responsivity of the detector 
on the history of flux the detector has been exposed to.) Although 
the response of the MIPS Ge detectors is frequently measured 
using flashes (``stims'') from a light source within 
the camera, drift and memory effects are not perfectly corrected. 
In scanning or rapid chopping observations, the rapid appearance 
and passing of the sources in a given detector pixel limits drift 
and memory effects to those associated with short time constants. 
The stims do a good job of tracking the detector response on 
short time scales, and the Spitzer Science Center (SSC) reduction 
does a good job of producing repeatable point source fluxes. 
The Pipeline reduction procedures are discussed in detail in 
Gordon et al. (2005; but note that the implementation of these 
procedures by the SSC may differ in small ways). The Spitzer 
Data Handbook states that MIPS 160 \ensuremath{\mu}m fluxes are repeatable 
to about 4\% with an absolute uncertainty of about 20\%. However, 
these numbers do not apply to background fluxes; the instrument 
is not at all optimized for background observations where long 
time constant effects (``slow response behavior'') 
can complicate the reduction and interpretation of the data. 
According to the Data Handbook, point source measurements should 
be made by subtracting the median background from each frame 
before analysis, i.e. the background information should be discarded. 
The Data Handbook also gives instructions for reducing observations 
of bright extended emission sources for Ge detectors; the instructions 
for reducing observations of faint extended structures are conspicuously 
absent. This underscores the point that MIPS is not optimized 
for background measurements. The SSC has not thoroughly investigated 
the properties of the instrument for low-flux very extended structures, 
the fluctuations in the sky background. Nonetheless, we shall 
demonstrate below that the problems associated with our data 
are manageable, and a great deal of information is available 
in these maps. 

\subsection*{2.3 Map Reduction}

Our co-added maps, which have pixels of the nominal camera pixel 
size, give the average of the flux measurements closest to each 
pixel center. No re-sampling of the maps and no distortion corrections 
have been applied at this time because we are interested in structures 
much larger than the camera pixel size. 


\subsubsection*{2.3.1 Data Selection}

This work includes data from pipeline version 11, which includes previously embargoed data
 \footnote{In our initial reductions and pre-publication versions of this 
paper, the data were processed by the SSC pipeline version 10. 
The version 11 data show improved quality.}.  Some data covering our selected 
fields was excluded from our analysis due to data quality or instrument settings.
In the SLH maps, a rectangular patch of sky near (\ensuremath{\alpha},\ensuremath{\delta}) =
 (163.5\ensuremath{^\circ}, 57.6\ensuremath{^\circ}) is covered in AORs 
 9632512 and 9832256.  These data have 
high noise and so were not used in our map, leaving a small rectangular 
region without data, which we refer to as the ``window''.   These data sets clearly 
dominated our rms noise map, described 
below, with a median rms of 0.33 MJy/Sr in these regions vs. 
0.23 MJy/Sr typical of the rest of the map. (The noise had structure 
in the scan direction, and so is likely due to drift or memory 
effects; it is not random noise.) We also excluded data from 
PID81, which were taken with different instrument settings (including 
stim period) and were not appropriate to combine with the rest 
of our data. Table 2 gives a list of AORs which identify the data 
used. In the FLS map, all data were used. 

\begin{table}
\begin{minipage}[t]{\columnwidth}
\caption{SLH Data and Zero Point Constants }
\label{table_offsets}
\centering
\renewcommand{\footnoterule}{}  
\begin{tabular}{cc}
\hline\hline
AOR Key & Offset \\
\hline

{\raggedright {\small 5177088}} & 
{\raggedright {\small -0.0410045}}\\

{\raggedright {\small 5177344}} & 
{\raggedright {\small 0.0838553}}\\

{\raggedright {\small 5179136}} & 
{\raggedright {\small -0.0582686}}\\

{\raggedright {\small 5179392}} & 
{\raggedright {\small 0.0984183}}\\

{\raggedright {\small 5179648}} & 
{\raggedright {\small 0.0494460}}\\

{\raggedright {\small 5179904}} & 
{\raggedright {\small -0.0894248}}\\

{\raggedright {\small 5180160}} & 
{\raggedright {\small 0.0509532}}\\

{\raggedright {\small 5180416}} & 
{\raggedright {\small -0.00977480}}\\

{\raggedright {\small 5180672}} & 
{\raggedright {\small -0.0789033}}\\

{\raggedright {\small 5180928}} & 
{\raggedright {\small 0.0552630}}\\

{\raggedright {\small 5181184}} & 
{\raggedright {\small 0.0140596}}\\

{\raggedright {\small 5181440}} & 
{\raggedright {\small -0.0669587}}\\

{\raggedright {\small 5184768}} & 
{\raggedright {\small -0.0396491}}\\

{\raggedright {\small 5185024}} & 
{\raggedright {\small -0.143126}}\\

{\raggedright {\small 6592512}} & 
{\raggedright {\small 0.00856844}}\\

{\raggedright {\small 6592768}} & 
{\raggedright {\small -0.124137}}\\

{\raggedright {\small 6593536}} & 
{\raggedright {\small 0.0775429}}\\

{\raggedright {\small 6593792}} & 
{\raggedright {\small -0.0734718}}\\

{\raggedright {\small 6594048}} & 
{\raggedright {\small -0.0432453}}\\

{\raggedright {\small 6594304}} & 
{\raggedright {\small 0.00904450}}\\

{\raggedright {\small 6595072}} & 
{\raggedright {\small 0.0809884}}\\

{\raggedright {\small 6595328}} & 
{\raggedright {\small -0.0681358}}\\

{\raggedright {\small 6596096}} & 
{\raggedright {\small -0.0501895}}\\

{\raggedright {\small 6596352}} & 
{\raggedright {\small -0.0108055}}\\

{\raggedright {\small 7770368\footnote{These two AORs are from the validation scans, taken in 
a different epoch from the rest of the data, when the estimated 
zodical light contribution was 0.22 MJy/Sr lower.}}} & 
{\raggedright {\small 0.463996}}\\

{\raggedright {\small 7770624$^b$\label{These two AORs are from the validation scans, taken in 
a different epoch from the rest of the data, when the estimated 
zodical light contribution was 0.22 MJy/Sr lower.}}} & 
{\raggedright {\small 0.331227}}\\

{\raggedright {\small 9628672}} & 
{\raggedright {\small 0.0402081}}\\

{\raggedright {\small 9628928}} & 
{\raggedright {\small 0.0327531}}\\

{\raggedright {\small 9629440}} & 
{\raggedright {\small -0.0725303}}\\

{\raggedright {\small 9629952}} & 
{\raggedright {\small 0.0269623}}\\

{\raggedright {\small 9630208}} & 
{\raggedright {\small -0.0331634}}\\

{\raggedright {\small 9630464}} & 
{\raggedright {\small -0.0134986}}\\

{\raggedright {\small 9630720}} & 
{\raggedright {\small -0.0330788}}\\

{\raggedright {\small 9630976}} & 
{\raggedright {\small -0.000532301}}\\

{\raggedright {\small 9631744}} & 
{\raggedright {\small 0.0146916}}\\

{\raggedright {\small 9632000}} & 
{\raggedright {\small -0.0804646}}\\

{\raggedright {\small 9633280}} & 
{\raggedright {\small -0.0644805}}\\

{\raggedright {\small 9633536}} & 
{\raggedright {\small 0.0322532}}\\

{\raggedright {\small 9633792}} & 
{\raggedright {\small -0.0983279}}\\

{\raggedright {\small 9634048}} & 
{\raggedright {\small 0.119130}}\\

{\raggedright {\small 9634304}} & 
{\raggedright {\small -0.00922350}}\\

{\raggedright {\small 9634560}} & 
{\raggedright {\small 0.0412873}}\\

{\raggedright {\small 9634816}} & 
{\raggedright {\small 0.00126636}}\\

{\raggedright {\small 9635072}} & 
{\raggedright {\small -0.0397363}}\\
\hline
\end{tabular}
\end{minipage}
\end{table}

{\subsubsection*{2.3.2 Stim Correction}}

In both ``raw'' maps (direct co-adds of Basic Calibrated 
Data, or BCD), sharp parallel lines can be seen perpendicular 
to the scan pattern (See Figure 2). These correspond to the frames 
taken just after the stim flashes; the effect is referred to 
as a ``stim-flash latent''. The stim flash itself 
necessarily contributes to memory effects in the detectors, as 
would any illumination. These effects depend on the integrated 
illumination history, not just the instantaneous brightness. 
Although the flash is very bright compared to typical source 
and background fluxes, it is very short in duration, producing 
small integrated fluence, and so causes only a small memory effect. 
Unfortunately, the small stim latent is significant compared 
to the faint CFIB.

Figure 3 shows BCD light curves from the observations of the 
SLH field at 160 \ensuremath{\mu}m (AOR 5177344) folded at the stim 
period. Here, each period of data was divided by the median of 
the data during that period in order to remove variations of 
the sky level during the observation. In most pixels, the residual 
effect is 10\% - 15 \% in the time bin or DCE after the stim flash, 
much greater than the error bars from the variation in sky flux 
(3-5\%; see Figure 3). In the third and following pixels after 
the flash, however, the effect (0-3\%) is rarely statistically 
significant. We applied a correction for this stim flash latent 
residual, simply the inverse of the folded and normalized timelines. 
A comparison of the sky maps made with and without the stim flash 
latent corrections, shown in Figure 2, is dramatic. The dominant 
structure in the raw map, the lines made by the stim-flash latents, 
has been removed.

{\subsubsection*{2.3.3 Illumination correction}}

The Spitzer Data Handbook recommends that for extended source 
observations an illumination correction be made. Illumination 
corrections were made similar to the technique for making CCD 
flats: We found the median value for each detector over a large
set of data. This median ``image'' was then normalized, 
and all data in the sample were then divided by the median image. 
We tested making a different illumination correction for each 
scan (i.e. correcting only the data between each change of scan 
direction, as recommended in the Data Handbook), and also using 
the same correction for an entire AOR. If an illumination correction 
were beneficial, we would expect that the rms deviation between 
repeated measurements of the same sky would decrease. In both 
cases, no significant decrease in rms deviation was achieved.

{\subsubsection*{2.3.4 Zero-Point Errors}}

In our initial co-added maps, regions associated with a given 
AOR appeared to have discontinuous flux on the borders of regions 
covered by other AORs. On further investigation, we found that 
this was reflected in a systematic discrepancy between measurements 
of the same sky position during different AORs. This can easily 
be seen in the histogram of median AOR data-map deviation. (We 
define the deviation \ensuremath{\delta}$_{i}$ of each measurement \textit{f}$_{i}$ in a 
given AOR at a sky location \textbf{r}(i) to be the difference between 
that measurement and the map at the location of the measurement, 
or \ensuremath{\delta}$_{i}$= \textit{f}$_{i}$(\textbf{r}(i)) - map(\textbf{r}(i)). 
The map value is just the average of all flux measurements \textit{f} at any given 
postition, map(\textbf{r}(i)) = \texttt{<}\textit{f}(\textbf{r}(i))\texttt{>}. The median 
deviation for a given AOR is then D = median(\ensuremath{\delta}$_{i}$) for 
all i in the AOR.) The histogram of median AOR deviation is shown 
in Figure 4 for our selected data. The obvious outliers in the 
SLH data are from observations of the ``validation
region'' of the survey. The validation region data were taken 2003 
December 9, long before the rest of the survey, 2004 May 4-9. 
According to the SSC tool SPOT, the level of zodiacal light is 
estimated to be 0.22 MJy/Sr lower during the validation observations. 
(Zodiacal light is that scattered by dust within our solar system, 
a ``local'' effect that changes with the Earth's, 
or the nearby Spitzer Observatory's position in solar orbit.) 
However, the deviation in the other data sets taken at the same 
time is apparently instrumental in nature.

We used numerical techniques to find the minimum variance (of 
repeated observations of the same sky pixel) set of additive 
zero point constants to reduce this problem.  The set of minimum variance constants 
is given in Table 2 for our SLH map.

{\subsubsection*{2.3.5 Zodical Light}}

No zodiacal light correction was applied (except in the correction 
for a different observation epoch, as described above); we show 
in Section 3, Map Power Spectra, that this produces 
negligible effects on our results.

{\subsubsection*{2.3.6 Full Maps}}

The full SLH map with all the corrections described thus far 
is shown in Figure 5. The FLS map is shown in Figure 6.

{\subsubsection*{2.3.7 Striping}}

The map has structure which clearly follows the scan pattern; 
such structure is referred to as ``stripes''. This 
can be seen easily on a computer screen, though it often shows 
up poorly in printed reproductions. The scans are all approximately 
parallel in the maps (except for small validation regions), enhancing 
this structure. In Figure 7 we show the SLH map smoothed at a 
size of 1/2 the instrument width to show this effect more clearly; 
the figure has been rotated so the scans are vertical, and the 
structure is also vertical. The structure in the map looks like 
large blocks, the regions covered by each AOR, rather than thin 
stripes. 

Monotonic drift in most detectors in the array, zero-point offsets 
in most detectors, and memory effects can contribute to stripes. 
We compared the individual detector timelines to the average 
of all measurements at the same map points, and determined that 
there was no significant monotonic drift in most pixels. However, 
we showed above that a median deviation with the same characteristics 
as a residual zero-point offset is present in the data. This 
would cause structure at AOR boundaries, as is observed. Detector 
memory effects are extremely difficult to test for, however, 
and the contribution of this effect remains unknown. Detector 
memory effects can cause bright extended emission regions to 
appear ``smeared'' across the map in the scan direction, 
resembling stripes. When the instrument finishes scanning a region 
and points to a new one, the detector is usually annealed, and 
its flux history is ``reset''. The background in 
the new region will reflect the detector's new history while 
in that region, which will be different than in surrounding regions; 
hence the discontinuities at the boundaries of the regions covered 
in each AOR could be caused by memory effects. There is one region 
where significantly less striping occurs in the SLH map, in the 
validation region (see Figure 5 and Figure 7), where some scans 
were done at a large angles to others. (Such scans are said to 
be ``cross-linking'', described below.) The worst 
striping is at the extreme right edge of the map in Figure 5 and Figure 7, 
where several of the brightest, more extended emission regions 
are located. This region was not included in our power spectrum 
analysis below.

\section*{3. Power Spectrum Analysis}

\subsection*{3. 1 Power Spectrum Calculations}

Our analysis closely follows that of Lagache et al. (2000). The 
basic steps for analysis of the CFIB are: (1) A power spectrum 
is made from the map. (2) Noise is subtracted from the power spectrum, and 
it is corrected for instrumental response. (3) 
The local foregrounds are then subtracted to yield 
the power spectrum of the CFIB. This section will cover all steps 
of this analysis, and the results will be covered in the following 
section. 

We analyze structure using a simple two-dimensional discrete 
Fourier Transform on square map sub-sections (shown in Figure 5 
for the SLH map). We report only the average magnitude squared 
of the Fourier components P(\textit{k}) in binned \textit{k} intervals 
(\textit{k} =(k$_{x}$$^{2}$ +k$_{y}$$^{2}$)$^{1/2}$). 
We did not apodize our maps prior to 
power spectrum analysis in order to preserve the information 
in the corners.


 In order to measure the noise power spectrum, two separate maps 
were made from the alternating (i.e. even and odd) measurements 
at each sky location. The even and odd maps were then subtracted 
to make a difference map which was analyzed to determine the 
noise power spectrum. 

\subsection*{3.2 Map Preparations}

{3.2.1 Missing Data}

Certain artifacts of the maps had to be ``repaired'' 
before proceeding to power spectrum analysis. Our SLH field map 
contains an approximately rectangular section, the NW corner 
of the ``window'', 0.29\ensuremath{^\circ} \ensuremath{\times} 0.63\ensuremath{^\circ} 
(66 \ensuremath{\times} 143 pixels) for which all AORs have excessively 
high noise (see section 2.3.1). Because the area takes up only 
a small fraction of the map area (0.18 deg$^{2}$, 2.1\% of our square 
map size), the loss of these data should have only a negligible 
effect on our results. We compared several methods for replacement 
of the data in this region: replacement with contiguous sections 
of data taken ``above'' and ``below'' 
the window, from both ``sides'' of the window, and 
finally we replaced the data with a smooth fifth order polynomial 
fit to the data around the window. The power spectrum results 
were insensitive to the choice of replacement method (or interpolation 
order), and we finally adopted the smooth fit.


There are unobserved pixels in the maps; in the FLS about 1.3 
\%, in the SLH 3 \ensuremath{\times} 10$^{-4}$ of the area in our square map 
was unobserved (in addition to the missing rectangle). We replaced 
all small groups of unobserved pixels with a local median of 
non-zero pixel values with a center-to-center distance of \ensuremath{\leq} 
10 pixel widths. These replacements had minimal effect on the 
final power spectrum.

{3.2.2 Source Removal}

We removed sources from our maps before calculating power spectra 
using the SEXTRACTOR program (Bertin \& Arnouts \cite{Bertin}) to find 
source locations (x,y) and x and y sizes (s$_{x}$, s$_{y}$). Sources 
with isophotes more than 1.5 \ensuremath{\sigma} (0.53 MJy/Sr) above the 
local background were removed, if they had more than four or 
more adjacent pixels above the threshold. At the location of 
each source centroid, a circular region within diameter d$_{replace}$  
was replaced with local background values. The resulting one-dimensional 
source sizes s = maximum (s$_{x}$, s$_{y}$), yielded good results in 
the range of about 5.65 to 9.5 pix. For s \texttt{<} 5.65 pix we 
used d$_{replace}$ = 5.65 pix. For 5.65 pix \texttt{<} s \texttt{<} 9.5 pix 
we used d$_{replace}$ = s. For s \texttt{>} 9.5 pix we used a log truncation, 
d$_{replace}$ = 9.5 + 1.5 ln(s/2-3.75), then we truncated that result 
to enforce a maximum of 14 pix.

For each pixel in the replacement region, the median of an annulus 
of inner and outer diameters 3 d$_{replace}$ and 4 d$_{replace}$ replaced 
the pixel value. In the end, the power spectra were extremely 
insensitive to the widest possible range of d$_{replace}$, detection 
threshold, etc., except at the highest frequencies, which have 
no effect on our conclusions.

{3.2.3 Sub-Sample}

Because our power spectrum analysis software only accepts square 
maps, and because we wished to analyze the largest possible contiguous 
areas we analyzed the square sub-samples of each map indicated in Figures 5 and 6. 
We excluded a region of the far right edge in Figure 5 from our 
SLH map because of high cirrus and probable memory effects. The SLH
sub-sample area is 8.47 deg$^{2}$.

\subsection*{
3.3 Features of the Raw Power Spectrum}

The power spectra of the SLH and FLS maps are shown in Figure 8 
and Figure 9. The following features are of interest: The lowest 
frequency bins look like a simple power law; in this region IR cirrus emission is known to dominate.  
In the mid-frequencies, there is excess emission above this power law.  This is the signal due to cosmological sources.   At the high frequency end, the signal 
is strongly modulated by the instrumental response function,
the power spectrum of the point spread function (PSF) of the 
instrument and telescope. The power spectrum of the PSF provided 
by the SSC (simulated by the STINYTIM routine) is shown in Figure 10. 
At the highest frequencies, the signal becomes dominated by noise; noise power is within a factor of two of signal by  \textit{k} $>$ 0.7 arc min$^{-1}$.

\subsection*{
3.4 Systematic Effects}

{\subsubsection*{3.4.1 Zodical Light}}

The zodiacal light contribution at 160 \ensuremath{\mu}m is small compared 
to other wavelenghts, and so we did not remove zodiacal light 
from our maps prior to analysis. At 160 \ensuremath{\mu}m, a simple 
smoothed planar fit to the zodiacal background values predicted 
by SPOT was analyzed to determine the effect on the power spectrum. 
We estimate that zodiacal light contributes less than 2.5\% of 
the spectral power in any bin, with a maximum contribution at 
low \textit{k}.

{\subsubsection*{3.4.2 Structure Due to Zero-Point Offsets}}

As shown in section 2.3.4, there are systematic deviations in 
each AOR data set that have the character of a zero-point offset. 
Such a set of zero-point offsets might cause false structure 
because the scan pattern is rather regular. We tested the effects 
of such offsets on the power spectrum by producing a known, ``synthetic 
sky'', then simulating observation of  this ``synthetic sky'', including the addition of noise and offsets.  
 Comparison of the known input synthetic sky and the 
resulting maps will then give the systematic effects due to the 
zero-point offsets. 

We produced the synthetic sky with a very simple CFIB + single-component foreground 
cirrus model. We used actual cirrus images from ISSA plates 
 re-binned to the same pixel dimensions as in our map. This 
cirrus foreground has the desired -3 slope spectrum. We added 
a Poisson distributed synthetic CFIB signal (i.e. with a flat power spectrum), then 
scaled the two components to the power measured in our SLH map 
for the cirrus and CFIB, respectively. We then simulated scanning observations, using 
the same scanning pattern as in the SLH observations.

\textbf{Offset Correction Simulation Test:} In our first test, we essentially measure the ability of our mapmaking software to correct for offsets given the SLH scan pattern and measured noise.  We started with our synthetic sky  simulation and added the same noise power as observed in the 
SLH, and \textit{the same set of deviations measured in the actual BCD
data}.  We made different realizations of our simulated maps, adding 
the measured deviations from the SLH data to the simulated data 
sets, but assigning them to the different AOR data sets 
in a different, random order in each realization.  (We used different realizations in order to understand effects similar to offsets in a very general way.)  We reduced these simulations in the same way as our real maps, fitting for zero-point offset corrections.   At this point, most of the effects of offsets should have been corrected, and we expect little effect on the power spectrum. 

The simulations showed only a very small effect on the power spectrum ($< 15\%$ for $k< 1$ arc min$^{-1}$), negligible compared to the uncertainty due to the fit errors at low \textit{k}.  The software succeeded in fitting good offset corrections. The behavior was consistent among all our simulations.   

\textbf{Residual Deviation Simulation Test:} In this test we observe the effects of \textit{uncorrected} offsets on the power spectrum, to match those seen in Fig. 4b.  Starting again from synthetic sky maps, 
we added the same random noise power as observed in the 
SLH, and \textit{the same set of residual deviations measured in our actual final SLH map} (see Fig.  
4b). We made different realizations of our simulated maps, adding 
the same measured residual deviations , but assigning them to the different AOR data sets 
in a different, random order in each realization.  (We used different realizations in order to understand effects similar to the measured residual deviations in a very general way.) We reduced these simulations in the same way as our real maps, but \textit{without} correcting these offsets.  
This had a larger effect on the power spectrum,  ($< 25\%$ for $k< 1$ arc min$^{Ð1}$; See Fig. 
11), but still very small compared to the fit errors.  Below, we use this function to make a correction for this effect, the inverse of the function shown in the figure.  

\subsection*{
3.5 CFIB Analysis}

The CFIB analysis requires correction of the raw map power spectrum for instrumental effects and removal of foregrounds.

\subsubsection*{
3.5.1 Instrumental Response Function}

The sky map may be described as the convolution of the real sky 
and an instrumental response function plus noise. The power spectrum 
of the sky may therefore be derived from the instrumental power 
spectrum minus the noise power spectrum, divided by the response 
function of the instrument. In Lagache \& Puget  (\cite{LagachePuget00}) and Miville-Deschenes, 
Lagache, and Puget (2002), the instrumental function was derived 
from the power spectra of cirrus-dominated maps by assuming that 
the (cirrus-dominated) sky power spectrum was dominated by a 
-3 slope power law. Large Spitzer maps of such regions are not 
readily available. In the analysis below, we assume that the 
PSF dominates the instrumental response function, and approximate 
our instrumental response function by the power spectrum of the 
PSF given in Figure 10.


Since we are looking for departures from the smooth cirrus power 
law in our power spectra, assuming a smooth response function 
without justification could lead to the mis-identification of 
instrumental features as features of the sky. In order to address 
this problem, we used the limited cirrus observations available. 
The FLS galactic observations do include a 15' X 1\ensuremath{^\circ} 
strip observation of cirrus, AOR Key 4962304. To reach scales 
large enough to be of interest, we analyzed one-dimensional strips 
in the long direction of this map, reduced in the same way as 
our extragalactic field maps. The cirrus map spectrum (Fig.12)
shows a smooth power-law like function at the lowest map frequencies, 
and noise and artifacts dominate at high frequencies. Assuming 
that the cirrus is a perfect power law, and assuming this map 
is the convolution of this power law and the instrument function 
+ noise + artifacts (at high-frequency), we can draw an important 
conclusion from these data: from 0.1 to almost 0.01 arc min$^{-1}$ 
the instrumental function must be smooth, with bumps or wiggles 
\texttt{<}\texttt{<} factors of two.

\subsubsection*{
3.5.2 Residual Offset Corrections to the Power Spectrum}

We also applied a correction for our residual offsets (measured residual deviations) described in Section 3.4.2.  To correct for these residual offsets, we divided out the effect of these offsets in our simulations, the function given in Fig. 11. 

\subsubsection*{
3.5.3 Foreground Cirrus Subtraction}

Previous measurements of the sky on scales from arc seconds to 
much larger than our maps have shown that cirrus structure has 
a power law shape with a log slope very close to --3  (e.g., Wright 
\cite{Wright}, Herbstmeier et al. \cite{Herbstmeir}, Gautier et al. \cite{Gautier}, Kogut et 
al. \cite{Kogut}, Abergel et al. \cite{Abergel}, Falgarone et al. \cite{Falgarone}). This very 
steep slope means that this structure must dominate at the lowest 
frequencies (as shown in Figure 8). Following Lagache \& Puget  (\cite{LagachePuget00}), 
we subtract a power law fit of the low-frequency structure from 
our power spectrum in order to remove the cirrus contribution. 
We fit a power law function to the lowest four bins of \textit{k} 
in order to get a good fit of the cirrus structure in the range 
of \textit{k} where it dominates.

\subsubsection*{
3.5.4 CFIB Spectrum}

After subtraction of noise and correcting for instrumental response 
and residual offsets, we obtain the sky spectrum. The final step in our analysis 
then is to subtract the foreground cirrus contribution, which 
we assume in advance to be a power law. Given the extensive evidence 
for this behavior, we decided to empirically determine our fit 
errors from the deviation in the log of our data from the even-weighted 
power law fit. The fit errors dominate the uncertainty in our 
CFIB measurements at low-\textit{k}. The CFIB fluctuation power spectrum is shown for the SLH field 
in Fig. 13.

\section*{4. Results}

\subsection*{
4.1 The CFIB Fluctuations Measurement}

{\subsubsection*{4.1.1 The SLH Fluctuation Spectrum}}

The observed CFIB power spectrum is described in 
gross terms as showing high power at low \textit{k}, decaying rapidly 
to \ensuremath{\sim}0.1 arc min$^{-1}$, with a relatively flat region \ensuremath{\sim}0.15 
- 0.4 arc min$^{-1}$.   If the sources of the CFIB were distributed at random 
in space, a flat power spectrum would be expected. What is observed 
is clearly different.  This excess CFIB power at low \textit{k} has been 
identified as the signature of clustering of CFIB sources, discussed 
in the following section. The error bars on the lowest few \textit{k} 
values are large due to the finite uncertainty in the subtracted 
power law fit. However, the remaining points have quite reasonable 
uncertainties, and we concentrate on these in our discussion. 
At the large \textit{k} values, the noise becomes significant and the instrumental correction becomes 
very large (see Fig.8; the figure is
cut off at \textit{k} = 1.0 arc min$^{-1}$. )

{\subsubsection*{4.1.2 Comparison with Predictions}}

In Lagache, Dole \& Puget (\cite{LDP03}), a pre-Spitzer 
simulation of the FIR sky was described (and made publicly available). 
The simulation used a relatively simple model galaxy distribution 
(``normal'' spirals + starburst galaxies) and evolution 
constructed to be consistent with IR - mm source counts available 
at that time, with sources distributed at random and interstellar 
foreground cirrus. Perrotta et al. (2003) showed the effects 
of predicted clustering on background power spectra. This theoretical 
prediction is shown superimposed on the power spectrum of the 
Lagache, Dole \& Puget (\cite{LDP03}) simulation in Figure 14. The range in which our low-\textit{k} excess power appears above the relatively flat region (Fig. 13) is 
very similar to the range given by the prediction, i.e. significant 
excess power above the cirrus power law at \textit{k \ensuremath{\geq}} 0.03 
arc min$^{-1}$ making a rapid decrease to the relatively flat region 
at \ensuremath{\sim} 0.2 arc min$^{-1}$.

%
\begin{table}
\begin{minipage}[t]{\columnwidth}
\caption{Low-\textit{k} Power Law Slopes}
\label{table_powspecresults}
\centering
\renewcommand{\footnoterule}{}  
\begin{tabular}{cc}
\hline \hline
{\centering \textbf{Map and Sample}} & 
{\raggedright \textbf{Cirrus Power Law Slope}\footnote{Power law fit to first four points. }}\\
\hline
{\centering SLH,corrected} & 
{\raggedright -3.13 \ensuremath{\pm} 0.28}\\

{\centering SLH,uncorrected} & 
{\raggedright -3.10 \ensuremath{\pm} 0.25}\\

{\centering FLS} & 
{\raggedright --3.12 \ensuremath{\pm} 0.55}\\

\hline
\end{tabular}
\end{minipage}
\end{table}
%

{\subsubsection*{4.1.3 Likely Systematics}}


These results are from a relatively simple analysis which relies 
on the assumption that the SSC PSF simulation gives the correct 
instrumental response function. However, it is unlikely that 
an instrumental feature would produce our excess low-\textit{k} (large-scale) 
power\textit{.} Large and medium scale power in the instrumental response 
function can only come from changes in the instrument over times 
comparable to observation of the whole map; the camera PSF has 
no structure on large scales. The map was sampled rather uniformly 
on the large scales, with essentially the same scanning overlap, 
redundancy, scan speed, etc. across the map. The FLS cirrus strip 
power spectrum was quite smooth, showing that there can be no 
bumps in the instrumental response in this frequency range. The 
ISO instrument (Lagache \& Puget  \cite{LagachePuget00}) also had no features in 
the instrumental response in this frequency range, and in general, 
such features would not be expected. The most interesting features 
we discuss here are mostly at intermediate size scales, i.e. 
0.03-0.2 arc min$^{-1}$. We note that the map sampled several regions 
large enough to observe this behavior. We also note that spurious 
artifacts generally occur at much smaller spatial scales.


In section 3.4.2 we addressed the possibility that systematics 
with the character of an instrumental zero-point offset could 
cause erroneous structure in our power spectra. The result of 
our simulations showed that there was a small
effect, and we corrected for this. The consistency of our simulations 
makes us confident in this correction. In the end, however, the 
erroneous structure had essentially little effect on the results, 
producing almost no effect on the cirrus fit.  Table 3 
lists the power law fits to the SLH power spectrum with and without 
the correction, and they are essentially indistinguishable. Because 
the dominant error in the CFIB spectrum is the power law subtraction 
uncertainty at the mid-k points of interest, and because the 
power law fit did not change, the correction yielded no significant 
change in the results.  In all our offset simulations, effects on the 
power spectrum were small and had no significant effect on our results; 
our results are therefore robust against such systemtic effects, within our errors.

{\subsubsection*{4.1.4 Dispersion in the SLH Results}}

The SLH sub-sample was selected \textit{a 
pasteriori}, avoiding bright cirrus regions and map edges. Reducing 
a few different sub-sample regions yielded some larger uncertainties 
in the power law fits, with larger errors as the subsample region 
moves closer to the bright region at the right edge of the map. 
All power spectra are consistent, however. 

We note that the two SLH validation AORs yielded the largest 
deviation even after zero-point correction (Figure 4). Removal 
of these two AORs from the map yielded results consistent with 
those reported above.   Because we did not find obviously anomalous behavior in the region 
covered by these data in the difference maps, however, we judged 
that elimination of these data may not be justified, therefore  we did 
not eliminate these data.

{\subsubsection*{4.1.5 First Look Survey Results}}

The FLS field power spectrum has no evident turn-off from the 
cirrus power law until about 0.09 arc min$^{-1}$, whereas in the 
SLH power spectrum, the turnoff was at \ensuremath{\sim} 0.03 arc min$^{-1}$. 
The difference is likely due to the greater strength of the cirrus 
emission in the FLS field. The cirrus power is more than 10 times 
higher in the FLS at 10$^{-2}$ arc min$^{-1}$ than in the SLH. \\
The power law fit was again made from the first four bins. However, 
the field is smaller, so the first four bins are at higher frequency 
than in the SLH. In this field, the power law fit has larger 
uncertainty than in the SLH.  Because of this larger uncertainty, 
the signal in the low to mid-\textit{k} range of interest has large 
error bars, and the resulting CFIB measurement is not interesting. 
This result is not unexpected, however: the combination of strong 
cirrus in the field (requiring very small fit error for CFIB 
measurement), and sampling the cirrus at a higher \textit{k} (because 
the field is smaller) where the CFIB can interfere makes for 
an \textit{a priori} difficult measurement.

\subsection*{
4.2 Comparison to Previous Results}


The SSC provides a tool to separately predict the local, ISM, 
and extragalactic background components (part of the SPOT software) 
incorporating HI data and IR measurements (essentially the Schlegel, 
Finkbinder \& Davis (\cite{Schlegel}) Results), and knowledge of the MIPS 
instrument. Table 1 shows that in the SLH, the median total background 
is more than 18\% off the prediction. Since the predictions 
of SPOT agree very well with other measurements in this wavelength 
regime (e.g. ISO measurements in the FIRBACK fields; Lagache 
\& Dole 2001), this suggests that some modification of the MIPS calibration is in order. 
The MIPS 160 \ensuremath{\mu}m camera has a known, significant optical 
light leak. The MIPS calibration, now based on asteroids, might 
cause a significant error without correction for optical background 
flux.


Lagache \& Puget  (\cite{LagachePuget00}) fluctuation spectrum results are plotted 
along with our results (the diamond symbols) in Fig. 13.  (The 
values were taken directly from their Figure 3.)  The overall agreement is 
very good, given the bin-to-bin scatter, differences in instruments, and realistic measurement 
and calibration uncertainties.   

Lagache \& Puget  (\cite{LagachePuget00}) reported that their results 
were consistent with a constant level. However, if \textit{k} \texttt{>} 0.5 arc min$^{-1}$ is ignored, the 
Lagache measurements apparently reflect the same increase in 
power toward low frequency as observed here, at all except their 
very lowest frequency point. (For \textit{k} $>$ 0.5 arc min$^{-1}$ the 
correction for the PSF becomes large for both measurements, and may not be reliable.) 
 

\subsection*{
5. Possibilities for Future Improvements}


At this time, the problems in the map are dominated by scan-pattern 
related structure. In terms of data reduction, various frequency 
space filtering methods for removing stripes seem promising, 
and have been demonstrated on other types of maps (see, e.g., 
Miville-Desch\^{e}nes \& Lagache \cite{Deschenes05}). In addition, various algorithms 
are available for optimum statistical weighting of map data. 
We found, however, that unless the striping in the maps are significantly 
reduced, these algorithms do not produce good results.


We find that the greatest potential for improving these measurements 
is in acquiring new data on the SLH field in a manner appropriate 
for background observations. The MIPS observations are truly 
exceptional compared to typical background observations in the 
degree to which ``cross-linking'' scans were \textit{avoided}. 
Virtually all cosmic microwave/mm background experiments incorporate 
cross-linking in their scanning strategy. Such a strategy causes 
each sky pixel to be re-sampled along significantly different 
scanning paths on the sky. A simplified example of cross-linking 
scans would be a series of two rectilinear raster maps of a region 
oriented at 90\ensuremath{^\circ} to each other. Two sky positions measured 
along the same scan across the first map would be measured on 
different scans in the second map. Comparison of repeated observations 
on the same and different scans allows identification and measurement 
of any systematics that affect the measurements differently on 
the same and different scans. In the general case, this technique 
allows inter-comparison of measurements made close together in 
time (i.e. on the same scan) with those made at much longer time 
scales. In our case, this comparison would clearly identify and 
measure the zero-point offsets. In all of the large Spitzer surveys, 
rectangular regions of the sky were observed during each AOR, 
and then immediately repeated on almost precisely the same path 
(with only minor exceptions). The rectangular regions were oriented 
very nearly parallel, no scans were made along significantly different directions
 (except in the very limited verification observations), and these regions 
had only small edge overlap. Cross-linking was essentially \textit{minimized} 
in the existing surveys, permitting intercomparison among only 
a small fraction of data measured on different paths in different 
AORs. It seems clear that if cross-linking MIPS scans were added 
to existing Spitzer survey regions, significant improvement in the
background fluctuation measurement would result.

\section*{Conclusions}

In this paper we presented co-added maps from two large Spitzer survey 
fields observed with the 160 \ensuremath{\mu}m MIPS array.  Instrumental artifacts, and artifacts related to the scan pattern were observed, but these effects were substantially reduced by a variety of corrections.  We measured a cirrus 
power law slope of -3.13 \ensuremath{\pm} 0.28 in the Lockman Hole field. 
Subtracting this power law yielded a CFIB spectrum dropping rapidly 
from \textit{k} \ensuremath{\sim} 3 \ensuremath{\times} 10$^{-2}$ arc min$^{-1}$, to
to \textit{k} = 0.1 - 0.2 arc min$^{-1}$, and a flatter region at higher \textit{k}. 
Any assumption of a power law cirrus contribution plus a flat 
power spectrum from the ensemble of galaxies, i.e. from a random 
distribution of galaxy sources, is not consistent with our observations. 
The gross details of our results are consistent with predictions 
of a clustering ``signature'' in the CFIB power spectrum.

 \section*{Figure Captions}
 
 Fig. 1 Scan Patterns.  A simple representation of the scan pattern, 
a line connecting each pointing of the camera sequentially, is 
given above for both fields. Part a) shows the scan pattern for 
the FLS Extragalactic Field. Note that all scan paths are close 
to parallel, even those in the verification region (repeated 
observations denoted by denser region at center). Part b) shows 
the Lockman Hole (SLH) Scan Pattern. Note that all scan paths 
are close to parallel except for those in the validation region 
(repeated observations denoted by dense region of paths at approximately 
45 degrees to nearby paths, at upper middle of figure).

Fig. 2  Stim Latent Correction. The two figures show a small sub-section of the co-added SLH map.  The uncorrected sub-section, top, has regular, bright horizontal bands, perpendicular to the scan direction, which dominate the structure. These are due to the stim latent effect. The latent 
bands are completely removed by the correction process, shown 
in the lower figure. The intensity scale (MJy/Sr) is given at 
the right side of the figure; the spatial scale markings on the 
bottom and left sides of the figure are in units of instrumental 
pixel widths. The data are from the SWIRE Lockman Hole observations; 
but all data we examined have essentially the same effect.  

Fig. 3 Stim Latents. The data from three different 
detector pixels are shown folded at the stimulator flash period. 
To show possible time-dependence of the phenomenon, the first 
1/3 of the data were represented by a cross, the second third 
by a diamond, and the final third by a filled circle. The time 
bin after the stim flash (which occurs in bin 0 in the figure) 
is almost always high in almost all channels. (These data are 
from AOR 5177344.)     

Fig. 4  Median AOR Deviation.  In part a), above, the histogram of median deviation (median of data-map) for each AOR in the SLH map is shown.  The two outliers are 770624 and 770368, the validation region scans (-0.34, -0.48).  In part b), below, on the same horizontal scale, the histogram of residual median deviations is given (i.e. \textit{after} the optimal set of offsets was added to the data).  The added offsets make the distribution much more narrow.  The two outlying points are again AORs 770624 and 770368 (-0.08, -0.11). 
 
 Fig. 5 Full SWIRE Lockman Hole (SLH) Map.  The full SLH 160$\mu$m Map is shown above.  The small protrusions at the upper right are the edges of the validation region scans, which are roughly 
45\ensuremath{^\circ} from the main scans. The white rectangle indicates the square sub-region used in our analysis.  Regions further to the right are not analyzed here due to the strong cirrus emission in this region. The unnaturally smooth rectangle centered  near 250, 425 is the ``window'' referred to in the text with interpolated data (see text section 3.1).  The x and y spatial scales are in pix (15.9"/pix).  The bar at right indicates an intensity scale, in MJy/Sr.  The image is rotated so that the main scans are approximately vertical. 

Fig. 6 FLS Map. The full FLS map is shown in the picture above.  The white rectangle indicates the square sub-section analyzed in this paper.  The verification region, where additional observations were made, is a small region just below center.  The same spatial and intensity units are used as in the previous figure. The figure is rotated so that the scans are approximately horizontal. 
 
 Fig. 7 Lockman Hole Map Smoothed by 1/2 Detector Width.  This image of the map has been smoothed to better demonstrate the presence of large-scale structure in the scan direction.  Structure that looks  ``smeared'' in the scan (approximately vertical) direction is associated with the more extended and bright emission on the map, and  therefore could be due to memory effects. Note that in the validation region there is significantly less "smearing" than in the rest of the map due to sampling in different scan directions, i.e. cross-linking.  The same spatial and intensity units are used as in the previous figure.

Fig. 8 Power Spectrum of Lockman Hole Map.  Sources were removed from our square subsample of the SLH map for this power spectrum.  Filled circles indicate measured P(k), open circles indicate power corrected for instrumental response and residual zero-point effects.  The power law fit is shown as a dashed line in the range of measurements where the fit is made, and its extrapolation is denoted by a dotted line.  The continuous line at the bottom of the figure gives the measured noise spectrum.  See text for additional details. 

Fig. 9 FLS Map Power Spectrum.   The FLS power spectrum is shown in the figure above with the same symbols as in the previous figure.  Note that the much larger SLH field had measurements down to smaller k (Figure 8).  The power law extrapolation is not much higher than the data  up to 0.08 arc min$^{-1}$, and so the CFIB cannot be measured easily or accurately at low frequency in this field.  

Fig. 10 The Power Spectrum of PSF, generated by the routine Stiny Tim, is shown in the figure above. 

Fig. 11 Effect of Residual Offsets on the Power Spectrum.  The figure shows the relative effect of the observed residual offsets on the power spectrum.  See Section 3.4.2 for details.

Fig. 12 Cirrus Map Power Spectrum.  The figure shows the average of 8 power spectra each made from one dimensional slices of the FLS cirrus strip-map.  Each strip was the average of 5 contiguous lines of pixels along the long axis of the map. The low frequencies show a close fit to a power law, with deviations less than a factor of two.  Significant deviations would be expected simply due to the small sample size.  The close fit to a power law shows that the instrumental function must be smooth at  \textit{k}   \ensuremath{<}  0.1 arc min$^{-1}$.

Fig. 13 CFIB Spectrum from SWIRE Lockman Hole Map.  The figure shows our noise-subtracted, instrumental response, and residual deviation-corrected CFIB spectrum of the SWIRE Lockman Hole Map subsample (filled circles).  The low-frequency power appears to fall rapidly to \ensuremath{\sim}0.1 arc min$^{-1}$; there is a comparatively flat region \ensuremath{\sim} 0.15 - 0.4 arc min$^{-1}$.  At the high frequencies (where power drops rapidly) instrumental response corrections become large.  The error bars give the uncertainty in the cirrus power law subtraction and residual deviation correction only.  The cirrus subtraction uncertainty dominates at low frequencies.  The diamonds reproduce the results given in Lagache \& Puget  (\cite{LagachePuget00}). 

Fig. 14 Simulation and Clustering Prediction.  The power spectrum from the Dole et al. (2003) simulation (filled circles) shows a relatively flat Poisson component from  \textit{k} $>$ 0.05 arc min$^{-1}$, whereas the Perrotta et al. (2003) simulation (dashed line), which includes clustering, shows a significant decrease in power all the way out to  \textit{k} $>$ 0.2  arc min$^{-1}$.

\begin{acknowledgements}
 The authors wish to thank the staff and students of the Institute 
d'Astrophysique Spatiale (IAS), especially Guilaine Lagache, 
Fran\c{c}ois Boulanger, and Herv\'{e} Dole, for their kindness and 
collaboration during and after Grossan's time at the IAS. We 
wish to also thank the staff of the Spitzer SSC, especially at the Help 
Desk, the staff of Eureka Scientific, and the LBNL Institute for Nuclear and 
Particle Astrophysics. This work was supported by a Spitzer Archival Research grant, 
NASA grant 1263806, issued by JPL/Caltech.
\end{acknowledgements}

\end{document}